\pdfoutput=1
\documentclass[runningheads]{llncs}
\usepackage{hyperxmp}
\usepackage{hyperref}

\begin{document}

\title{If it's Provably Secure, It Probably Isn't:\texorpdfstring{\\}{ }Why Learning from Proof Failure is Hard}

\author{Ross Anderson\inst{1}, Nicholas Boucher\inst{2}}
\institute{Universities of Cambridge and Edinburgh\\
\and
University of Cambridge}

\hypersetup{
  pdftitle={If it's Provably Secure, It Probably Isn't: Why Learning from Proof Failure is Hard},
  pdfauthor={Ross Anderson and Nicholas Boucher}
}

\titlerunning{If it's Provably Secure, It Probably Isn't}

\maketitle

\begin{abstract}

In this paper we're going to explore the ways in which security proofs can fail, and their broader lessons for security engineering. To mention just one example, Larry Paulson proved the security of SSL/TLS using his theorem prover Isabelle in 1999, yet it's sprung multiple leaks since then, from timing attacks to Heartbleed. We will go through a number of other examples in the hope of elucidating general principles. Proofs can be irrelevant, they can be opaque, they can be misleading and they can even be wrong. So we can look to the philosophy of mathematics for illumination. But the problem is more general. What happens, for example, when we have a choice between relying on mathematics and on physics? The security proofs claimed for quantum cryptosystems based on entanglement raise some pointed questions and may engage the philosophy of physics. And then there's the other varieties of assurance; we will recall the reliance placed on FIPS-140 evaluations, which API attacks suggested may have been overblown. Where the defenders focus their assurance effort on a subsystem or a model that cannot capture the whole attack surface they may just tell the attacker where to focus their effort. However, we think it's deeper and broader than that. The models of proof and assurance on which we try to rely have a social aspect, which we can try to understand from other perspectives ranging from the philosophy or sociology of science to the psychology of shared attention. These perspectives suggest, in various ways, how the management of errors and exceptions may be particularly poor. They do not merely relate to failure modes that the designers failed to consider properly or at all; they also relate to failure modes that the designers (or perhaps the verifiers) did not want to consider for institutional and cultural reasons.

\end{abstract}

Security engineering has had a long and difficult relationship with `proof'. Some of the pioneers in our field were frankly dismissive: Donald Davies thought security proofs pointless, as you can prove a design resists the attacks you know of, but not the attack you don't know of yet. Others made serious contributions to the proof literature, notably the founder of this security protocols workshop, Roger Needham.

In a recent invited talk\footnote{LMU Munich, December 7 2022; \url{https://cast.itunes.uni-muenchen.de/clips/vj4LmcoIzH/vod/online.html}} the first author told the story of a number of security proofs that failed for various reasons:

\begin{enumerate}
    \item When the first author came to Cambridge to interview for a PhD place, Roger gave him a copy of his BAN logic paper~\cite{ban1990}. He used that to `prove' the security of UEPS, an electronic purse~\cite{ueps1992}. This got him a PhD place (and impressed his clients no end) but we later found a bug. The BAN logic let us verify that an electronic cheque was authenticated using an appropriate key, and was fresh enough. The bug was that we used two-key DES and the two keys weren't properly bound together. 
    \item The paper title was in the .sig of the Serpent coauthor Lars Knudsen. He'd proposed a block cipher provably secure against differential cryptanalysis~\cite{NK92}, but later found an easy attack on it of a different kind~\cite{JK97}.
    \item At Crypto 94, Mihir Bellare and Phil Rogaway produced a security `proof'~\cite{BR94} for optimal asymmetric encryption padding (OAEP) that caused everyone to start using it during the dotcom boom. Embarrassingly, Victor Shoup proved in 2001 that the alleged `proof' has a gap that cannot be fixed, but that it's probably OK by accident, but only for RSA~\cite{Shoup2002}.
    \item Larry Paulson proved the underlying SSL/TLS to be secure in 1999 in a paper that was highly cited, including when he was elected to the Royal Society 20 years later~\cite{Paulson99}. Yet TLS has been broken about once a year ever since. SPW attendees will be familiar with ciphersuite downgrades, timing attacks, the Bleichenbacher attack, the order of authentication and encryption, Heartbleed and much more. Larry's own view is `We still can’t prove stuff with sufficient granularity to track what happens in real systems'.
    \item Hugo Krawczyk produced another proof in 2001 which supported the `MAC-then-CBC' approach in some ciphersuites~\cite{Krawczyk2001}. In 2010, Kenny Paterson broke this CBC mode~\cite{MS13}.
    \item At CCS 2002, Mihir Bellare and colleagues proved that SSH's use of symmetric crypto was secure~\cite{ssh2002}. At Oakland 2009, Martin Albrecht, Kenny Paterson and Gaven Watson showed that it wasn't (SSH used an encrypted length field plus CBC, allowing cut-and-paste games, and the length field wasn't in the security model)~\cite{ssh2009}.
    \item The 4758 was evaluated to FIPS-140 level 4 and thus thought by the industry to be certified by the US government as `unhackable'. SPW veterans will recall the API attacks of the early 2000s where Mike Bond and Jolyon Clulow showed that however secure the device's hardware was, the software was anything but~\cite{1580505}. In fact, one of the attacks Mike found had been known to IBM and been dealt with by putting a `please don't do this' footnote in the manual. 
    \item Those security proofs offered for quantum cryptosystems that depend on entanglement make sense under some interpretations of quantum mechanics (Copenhagen) but not under others (de Broglie-Bohm, cellular automaton)~\cite{BA15}.
\end{enumerate}

Philosophers of mathematics and science have argued for years whether proof is Hilbertian, or partly social. See for example the Appel-Haken proof of the four-colour theorem, Bundy et al on persistent errors in proofs, and Shapin and Schaffer's `Leviathan and the Air Pump'. Where do we stand with security proofs?

A standard textbook notes that the definition of a `trusted' system depends on institutional factors. It can be:

\begin{itemize}
\item one I feel good about;
\item one that can break my security policy;
\item one that I can insure; or
\item one that won't get me fired when it breaks.
\end{itemize}

Larry's comments reflect a shift in the `verification' community over the past 25 years to describe their activity not as `proving' but as `modeling'. Mathematical models are not wrong unless they have mistakes in them; it's just that they are typically not complex enough to capture real systems.

This may be a fair defence for Larry, as some of the TLS attacks were on extensions that didn't exist in 1998, while others were on the crypto implementation which Larry abstracted. It is also, as SPW veterans will be well aware, a fair defence for Needham-Schroeder, whose 1970s paper assumed that all principals execute the protocol faithfully. By the 1990s, we had insider attacks, and suddenly neither the shared-key Needham-Schroeder nor the public-key version were secure.

The most extensive use of verification, since the Intel floating point bug, has been checking the correctness of CPU designs. Yet this entirely failed to anticipate the Spectre / Meltdown class of vulnerabilities, which exploit the timing consequences of microarchitectural state. Here, SAIL models say nothing about timing, so if you want to check the correctness of memory barriers, you have to do it entirely separately from your SAIL model. Writing a specification for a more granular model of the microarchitecture may simply be infeasible, as its whole point is to abstract away things that `don't matter'. So Spectre is out of scope for our current verification tools.

On the other hand, Kenny Paterson's two CBC attacks were held to disclose `mistakes' as cryptographers (as opposed to modelers) really should have known better. The same holds for Victor Shoup's demolition of the Bellare-Rogaway `proof'. 

Can the philosophy of mathematics say anything interesting? The Appel-Haken controversy flushed out an old dispute between Descartes' view that a proof should be capable of being held in the human mind, and Leibniz' view that a proof should involve a sequence of correct computations. In this sense, BAN is Cartesian while Isabelle is Leibnizian. The random-oracle security proofs fashionable in the 1990s were an interesting hybrid, in that when written down they were incomprehensible, except possibly to insiders; but when performed at the blackboard by Phil Rogaway or Mihir Bellare they appeared to make perfect sense. This odd mix of the Cartesian and the Leibnizian approaches has now fallen out of fashion, and the failure of OAEP may even have helped.

What about the philosophy of science? During the 1960s we learned from historians and philosophers of science such as Kuhn, Lakatos and Feyerabend that science is a social activity that gets stuck in a certain way of doing things -- a `paradigm', in Kuhn's terminology. When a paradigm runs out of road, it may take a scientific revolution to move to a better modus operandi. Well-known examples include the move from Aristotlean mechanics to the Galilean / Newtonian model; from phlogiston to oxygen in chemistry; and from Newtonian mechanics to quantum mechanics. Such major shifts involve a generational change, as younger scientists embrace the revolutionary new methods. Max Planck famously remarked that physics was advancing `one funeral at a time' as professors wedded to the old way of doing physics gave way to the next generation.

Entrenched communities of practice exist at much smaller scales than the whole of physics. The failures exposed by Paterson and Shoup occurred within the magisterium of the crypto community rather than on someone else's turf. They were mistakes because they were failures of `normal science', to use Kuhnian language. 

The 4758 failures were cross-community. Even though we gave IBM ten months' responsible disclosure they wasted it; the software folks at Yorktown were arguing with the hardware folks in Raleigh over whose fault it was. And however much the vendors patched their products, Visa kept breaking them again by standardising new transactions that were insecure, whether in combination with existing transactions or, in one case, even on their own~\cite{YAB+2007}.

A real eye-opener was the response of the quantum crypto crowd to dissent. If you remark to the quantum boys that you don't buy the story around entanglement, the initial reaction is amused condescension -- until they discover that you understand the mathematics and have thought hard about the Bell tests. Then there's fury at the heretic. Our first brush with this you can find by searching for the post entitled `collaborative refutation' on Scott Aaronson's blog~\cite{collab2013}. Unable to see what was wrong with the first paper that Robert Brady and the first author of this paper wrote on the subject, apart from the fact that it was clearly heretical, he invited his followers to engage in a pile-on. This underlined the arguments of historians and philosophers like Kuhn, Feyerabend and Schaffer: sociology matters in science, and even in physics. 
Further experience talking about `heretical physics' confirmed this in spades. 

Some might question the soundness of applying the full Kuhnian theory to tiny subdisciplines, such as the users of a particular verification tool or proof technique. No matter; many of the same insights may be drawn from Tomasello's psychological research on shared intentionality. One capability that humans have and the great apes lack is that we can develop goal-directed shared attention on a joint task; Tomasello argues that this was a key evolutionary innovation on the road to language and culture~\cite{Tomasello2014}. As he puts it:

\begin{quote}
   Thinking would seem to be a completely solitary activity. And so it is for other species. But for humans, thinking is like a jazz musician improvising a novel riff in the privacy of his room. It is a solitary activity all right, but on an instrument made by others for that general purpose, after years of playing with and learning from other practitioners, in a musical genre with a rich history of legendary riffs, for an imagined audience of jazz aficionados. Human thinking is individual improvisation embedded in a sociocultural matrix. 
\end{quote}

This perspective may give useful insight into why security proofs often fail. The genres within which they are developed are too restrictive. That is why error handling is often hard; errors often fall outside the genre. There are multiple barriers to dealing with them -- not just economic and institutional, but also cognitive and cultural. 

Such observations are not entirely new. In the 1930s, Upton Sinclair noted that it's difficult to teach anyone something when his job depends on not understanding it. However Tomasello's work opens the door for a modern social-science study of such phenomena, which may become ever more more important as supply chains become more complex and failure modes more recondite. 

This brings us to our final example, and the stimulus for writing this paper: the bidirectional coding vulnerabilities we recently discovered in both large language models and computer source code. The latter mostly got fixed while the former largely didn't~\cite{BA22}. This appears to have been largely cultural. Engineers who maintain kernels and compilers in C generally care about the patch cycle, while data scientists and NLP researchers who build deep neural networks generally don't. As machine-learning components end up in more and more systems, the ability to deal with errors may hinge on subtle interactions with human learning, developer cultures and institutional incentives.

\bigskip\noindent{\bf Acknowledgement:} We thank Sam Ainsworth for valuable discussions on verification of microarchitecture.

\bibliographystyle{splncs04}
\bibliography{sources}

\end{document}